\documentclass{pazh_engl}
\usepackage{graphicx}

\begin{document}

\title {\bf Broadband X-ray spectrum of XTE J1550-564 during 2003 outburst}
\author{\copyright 2004 
     V.A.Arefiev\inst{1}, M.G.Revnivtsev\inst{1,2},
     A.A.Lutovinov\inst{1}, R.A. Sunyaev \inst{1,2}
} 

\institute{
Space Research Institute, Moscow, Russia
\and
Max-Planck-Institut fuer Astrophysik, Garching, Germany
}

\date{April 8, 2004}
 
\authorrunning{AREFIEV ET AL.}
\titlerunning{BROADBAND X-RAY SPECTRUM OF XTE J1550-564 DURING 2003 OUTBURST}

\abstract{
Results of broadband INTEGRAL and RXTE observations of the Galactic 
microquasar XTE J1550-564 during outburst in spring 2003 are
presented. During the outburst the source was found in a canonical
low/hard spectral state.  
}
\maketitle

%

\section{Introduction}

X-ray transient source XTE J1550-564 was discovered with all sky
monitor ASM  onboard the RXTE observatory on September 7 1998 (Smith,
1998). Soon its optical 
and radio counterparts were identified by \cite{orz98}, \cite{campw98}.
During the outburst in 1998-1999 the source have demonstrated all
canonical spectral states of black hole binaries -- high/soft, low/hard and
intermediate. Spectral and temporal variability of the source was very 
complicated (\cite{homan98}). After two weeks of the source discovery
a giant X-ray flare was observed, when the source flux reached the 
value of 6.8 Crab (\cite{rem98}). Luminosity, that corresponds to this flux
was $9.4\times 10^{38}$~erg/s in the energy band 2-100 keV
(assuming the 
source distance 5.3 kpc, \cite{orz02}), that approximately equals to the
Eddington luminosity for this source. Recently CHANDRA observations
revealed weak X-ray emission of relativistic jets associated with the source,
which, according to their velocity and distance from the source, can 
be launched during its near-Eddington flare (\cite{toms03}).
Relativistic jets of the system, that also were observed in the radio
band (\cite{han01}), allowed to classify XTE J1550-564 as a microquasar. 
Optical observations demonstrated that the binary system has orbital
period 1.54 days and the mass of the compact object is $M\sim 10M_\odot$
(\cite{orz02}), that is significantly higher that it is supposed to be 
possible for a neutron star.

XTE J1550-564 demonstrated a set of strong outbursts: in 2000 
(\cite{toms01,rodr03}), 2001 (\cite{toms01b}), 2002 (\cite{bel02}) and 2003
(\cite{dub03}), however none of these outbursts were as complex and
powerful as the first
one (see Fig. 1). All subsequent outbursts had more or less standard
shape of the lightcurve, with fast rise and slow decay, and practically
had no spectral variability. It is very curious that the time
difference between consequent outbursts is approximately 1 year and stays
like that during last 6 years, while the source was never (at least 25 years)
observed before 1998.

In this paper we present results of analysis of observations of
XTE J1550-564 with INTEGRAL (Core program data) and RXTE observatories
(publicly available slew data) during its 2003 outburst.

\section{Data analysis and results}

International gamma-ray observatory INTEGRAL (Winkler et al. 2003a) was 
launched by a Russian rocket-launcher PROTON on Oct.17 2002 (Eismont et
al. 2003). 
The observatory have four instruments, that allow one to study cosmic X-ray
sources in X-ray, gamma and optical bands. As a part of the Core
program the INTEGRAL observatory performs systematic scans
of the Galactic plane searching for the X-ray transients and studying
the variability of detected sources (Winkler et al. 2003b).
During one of such scans in spring of 2003 a beginning
 of a new outburst of XTE J1550-564 was detected (\cite{dub03}). 

During INTEGRAL observations of the Galactic plane XTE~J1550-564 
was mainly out of the field of view of the JEM-X monitor
(except for TOO INTEGRAL observations of this source, which we do not
consider here), therefore we will use in this work the data of the
IBIS telescope  (detector ISGRI, Lebrun et al. 2003) and spectrometer
SPI (Vedrenne et al. 2003), that have large fields of view 
($29^\circ\times29^\circ$ É $35^\circ\times35^\circ$, respectively).

Data of IBIS/ISGRI were reduced using the method described
by Revnivtsev et al. (2004). Analysis of the SPI data
was performed with the help of standard package OSA 3.0 
\footnote{http://isdc.unige.ch/index.cgi?Soft+soft}. For the reconstruction of 
the source spectrum
from the data of IBIS/ISGRI we used the ratio of the source fluxes
in different energy bands to those of the Crab nebula, assuming
the spectrum of the Crab in the form
 $d N(E)=10 E^{-2.1} dE$ phot cm$^{-2}$ s$^{-1}$ keV$^{-1}$. 
In order to check the correctness of the applied algorithms and
to estimate the amplitude of systematic uncertainties we
studied a large set of Crab nebula observations. Results of this analysis
show that with the used software we still have 2-5\% systematic uncertainties
in the spectral reconstruction and $\sim$10\% in the absolute flux value
estimation. Analysis of SPI data with the OSA 3.0 package
systematically gives overestimated flux from the source, while provides
a correct spectral shape reconstruction. Therefore in the subsequent analysis
we will renorm the SPI flux values to those obtained with the help of
IBIS/ISGRI.
 
As the source was out of the field of view of JEM-X monitor in order
to complement the obtained hard X-ray information with the
standard X-ray band we analyzed publicly available data of the RXTE
observatory -- data of the all sky monitor ASM and slew parts of
observations of 
spectrometer PCA, performed in the period March 27 - April 21, 2003.
Data of the RXTE/PCA were reduced with standard tasks of the
LHEASOFT/FTOOLS 5.3 
package. In order to reduce systematic uncertainties in the output spectra
we limited ourselves with the data of only upper anode layers of the PCA 
detectors and excluded data of PCU0 detector from our analysis, 
because since 2000 it lacks propane veto layer.

\subsection*{Lightcurve}

INTEGRAL observed XTE J1550-564 several times during the spring of 2003
performing the Galactic plane scans. From the beginning of March 2003
and to the first detection of XTE J1550-564 on March 24 the source
position was observed by INTEGRAL/IBIS during $\sim$300 ksec.
An upper limit on the averaged source flux during this time
is approximately 1-2 mCrab in the 18-60 keV energy band.

The outburst, first detected by the IBIS telescope, started on March 24.2,
2003  
(\cite{dub03}). Then source several times was in the field of view of
INTEGRAL during the rising phase of the outburst and at the peak 
of its lightcurve (March 24-25 and April 8,2003). The X-ray light curve
obtained by INTEGRAL and RXTE is presented in Fig.2.
It is seen that the overall length of the outburst in the standard
X-ray energy band (RXTE data) equals approximately 50 days, the shape
of the light curve is slightly asymmetric with the raise phase $\sim$10 
days, and more smooth decay on a time scale of $\sim 35-40$ days.
Because of limited amount of available INTEGRAL data we can not
determine the length of the outburst in the hard X-ray and gamma energy
bands. 
However, we can note that the maximal flux of the source detected
in the soft/standard X-ray band ($\sim 70$ mCrab) is significantly
lower than that seen in the hard X-ray band  ($\sim 200$ mCrab), that 
indicates that the source had a hard spectrum.

\subsection*{Spectrum} 
Preliminary analysis of spectra of XTE J1550-564
showed that during the outburst the source demonstrated only
subtle spectral variability. Because of that in order to improve statistics 
we will use spectrum of the source averaged over all observations.

The Broadband averaged spectrum of XTE~J1550-564, obtained by INTEGRAL and 
RXTE
observatories during the source outburst in 2003 is presented in Fig.3.
Also for comparison we present the spectra of the source averaged over
hard state periods of its outbursts in 2000 (see e.g.\cite{rodr03}) and 2001.
It is seen that spectra of 2001 and 2003 are significantly harder that 
that of 2000.

Standard model of the accretion flow in the black hole binaries
in their low/hard spectral states considers that the optically thick and
geometrically thin accretion disk ends (\cite{pkr97}), 
for example evaporates (\cite{meyer00}), at distances of the order of
$\sim 10-100 R_g$ from the black hole and most of the energy release occurs
in the optically thin hot plasma cloud at $R<10-100 R_g$. X-ray
spectrum originates as a result of the comptonization of photons
in this hot region (see e.g. \cite{st80}). Part of this
hard spectrum can be reflected from outer optically thick cold
accretion disk (\cite{basko74}, \cite{gf91}).
For the spectral approximation of obtained data we used the comptonization 
model of Poutanen\&Svensson (1996) (model $compps$ in the spectral 
package XSPEC), which includes the reflection
from the cold medium.
This model describes the cutoff of the comptonized spectrum
at high energies more correctly than simple analytic approximation in the 
form ($F\propto E^{-\Gamma} \exp(-E/kT)$). Temperature of the seed 
photons was fixed at the value of  0.1 keV. Best fit parameters of
the applied model are presented in Table 1. Note that the obtained
optical depth of the comptonizing cloud is rather high $\tau \approx 4-5$, 
that is not very typical for accreting black holes in the hard state.
However, it can indicate that due to possibly high inclination of the
system we are looking to the central source through whole depth
of the hot cloud.

In the spectral approximation we assumed systematic uncertainties
at the level of 1\% and 5\% at every energy channel for RXTE/PCA and
INTEGRAL/IBIS respectively. An absorption column at the
line of sight 
to the source was fixed at the value $N_H=10^{22}$ cm$^{-2}$ determined 
by CHANDRA (\cite{kaar03}). 

It is necessary to note that the present uncertainties in the
crosscalibration of the RXTE/PCA and INTEGRAL/IBIS+SPI influence on the best
fit 
parameter of the reflection ($R=\Omega/2\pi$). Therefore, in spite of
statistically significant detection of the reflected component
in the spectrum of XTE J1550-564 this value should be treated with care.

An absolute width of the Fe fluorescent line at $E\sim6.4$ keV was
fixed at the value of $\sigma=0.1$ keV (undetectable width for the RXTE/PCA)
for all spectra, except for that of 2000 outburst. In this spectrum
the width was determined to be $\sigma\sim 0.6-0.7$ keV.


\section{Discussion}

The X-ray transient XTE J1550-564 is located in the Galactic plane
approximately $4^\circ$ away from well known bright X-ray source Cir~X-1
in the region that was observed by different 
instruments/observatories. After at least 25 years of the off-state the
source 
entered to the new phase, in which it generates outbursts approximately
every year. Such a behavior is drastically differ
from usual behavior of black hole X-ray transients \cite{mckl03}).

During two first outbursts (1998-1999 and 2000) the source demonstrated
numerous and complicated spectral transitions (it is especially 
applicable to the first outburst). During the first outburst (Fig.1, 
MJD 51036-51299) the source was observed in the soft/high spectral state
and demonstrated $M$-like light curve not only in the standard X-ray
energy band but also in the broad (2-200 keV) band (\cite{sobc00}, 
\cite{alex02}). Sobczak et al. (2000) described the spectrum of the source
in this state as a sum of two components -- a multicolor optically thick 
accretion disk (Shakura\&Sunyaev 1973) and a power law. In the first 
phase of the outburst  (before $\sim$ MJD 51150) the power law component
dominated, while later the disk component contributed the most to the
observed X-ray emission.

Second outburst (2000) was less violent, the source flux reached the 
value 1 Crab
in the energy band of the RXTE/ASM , and the source behavior was much 
simpler:
at the beginning and at the end of the outburst when the source X-ray 
flux was low 
it demonstrated the hard spectrum (Rodriguez et al. 2003). At the 
maximum of the
outburst the source changed its state into the intermediate, when 
significant
contribution from the accretion disk was visible.

Rodriguez et al. (2003) have found that during 2000 outburst the 
transitions between 
low and intermediate states have demonstrated hysteresis effect. 
First transition
from low to intermediate state happened at the flux level $\sim2.3
\times 10^{-8}$ ergs cm$^{-2}$ s$^{-1}$ in the energy band 2-200 keV, while
the back transition from intermediate to low state happened  at 
$\sim1.1 \times 10^{-8}$ ergs
cm$^{-2}$ s$^{-1}$. Indications on such hysteresis effect can be also 
found in the
first outburst. As it was found in the work of Sobszack et al. (2000) 
during first two 
days of the outburst of 1998-1999 the photon index of the source spectrum 
was less  
than 2 while the source flux was lower than few$\times10^{-8}$ erg cm$^{-2}$ 
s$^{-1}$. Contribution of the soft disk component at that time was not very 
significant (around 5\% of the total flux). A comparison of the data of 
RXTE/PCA and CGRO/BATSE (Alexandrovitch \& Arefiev 2002) support the
assumption that the
source during MJD 51063-51066 was in the low/hard state. After the 
transition to the
high state the source have never returned to the low state even when its
flux dropped below the detection limit of the RXTE/PCA ($\times10^{-11}$ erg 
cm$^{-2}$ s$^{-1}$).

During none of the following outbursts 2001-2003 the source flux 
reached the values when the transition to the high/soft state occurs (see
Table 1 and Belloni et al. 2002). The 2003 outburst was observed from its
first days and the source spectrum remained hard during the whole outburst.

For the spectral approximation of the data obtained during the 2003
outburst we used the comptonization model of Poutanen \& Svensson (1996). 
The best fit parameters of the model, applied to the spectrum of 
XTE J1550-564 averaged over all available data are presented in Table 1. 
For comparison in the Table 1 we also present best fit parameters of the 
same model, applied to the averaged hard state spectra of XTE J1550-564
of outbursts 2000 and 2001. It is seen that the spectrum of the source 
during 2003 outburst is significantly harder than that of 2000 and more 
similar to that of 2001.

Finally we would like to note an interesting fact - every next outburst 
of XTE J1550-564 is weaker than the previous one (see Fig.1) and the source 
spectrum become harder and harder. It can be assumed that it is happening 
because of different physical conditions near the black hole: at the 
beginning of the 1998 outburst the black hole was fed by a very massive 
accretion disk, that was accumulated over large period of time,
then with every new outburst the surface density of the accretion disk 
diminishes, mass accretion rate become smaller and the outbursts become 
weaker and harder. Applying this assumption we can predict that the next 
outburst will be even weaker than the last one.


Authors thank E.M.Churazov for development of the IBIS data analysis 
algorithms and for the supplied software. The work was supported by 
MINPROMNAUKI (grant of President of Russian Federation NSH-2083.2003.2) 
and the program of Russian Academy of Sciences "Non stationary phenomena 
in astronomy". Authors thank Integral Science Data Center (ISDC, Vesoix, 
Swiss) and Russian Integral Data Center (Moscow, Russia).
The work is based on observations with INTEGRAL, an ESA project with
instruments and science data centre funded by ESA member states
(especially the PI countries: Denmark, France, Germany, Italy,
Switzerland, Spain), Czech Republic and Poland, and with the
participation of Russia and the USA. We have used the
data obtained from Archive of High Energy Astrophysics (HEASARC)
of Goddard Space Flight Center.

\pagebreak

\clearpage

\begin{table*} 
\begin{tabular}{lllllll} 
\hline
Outburst& $kT^a$, keV& $\tau^b$&R, $\Omega/2\pi^c$&  EW$_{\rm line}^d$,  
eV&Flux$^e$, erg/s/cm$^2$&$\chi2$/d.o.f.\\ 
\hline 
2003   & $50\pm 10$ & $5\pm 1$& $0.25\pm 0.13$&$120\pm 30$&$3.8\times  
10^{-10}$  $^f$&1.20\\ 
2001   & $63\pm 6$  & $3.8\pm 0.4$& $0.27\pm 0.08$&$96\pm  
25$&$5.5\times10^{-9}$&1.16\\ 
2000   & $49\pm 2$  & $3.3\pm 0.2$& $0.7\pm0.1$&$ 99\pm20$&$2.0\times  
10^{-8}$&1.6\\ 
\hline 
\end{tabular} 
\begin{list}{} 
\item  Inclination angle was fixed at $i=73^\circ$ level. (Orosz
 et al. 2002) 
\item $^a$ -- Temperature of coronal electrons. 
\item $^b$ -- Optical depth. 
\item $^c$ -- Geometry factor, representing the fraction of reflected component. 
\item $^d$ -- Equivalent width of the fluorescent Fe-line. 
\item $^e$ -- Flux at 3-200 keV. 
\item $^f$ -- Flux was calculated by normalization RXTE/PCA data to the
INTEGRAL/IBIS flux level. 
\end{list} 
\end{table*}

\clearpage

\begin{figure*}[t]
\includegraphics[width=14cm]{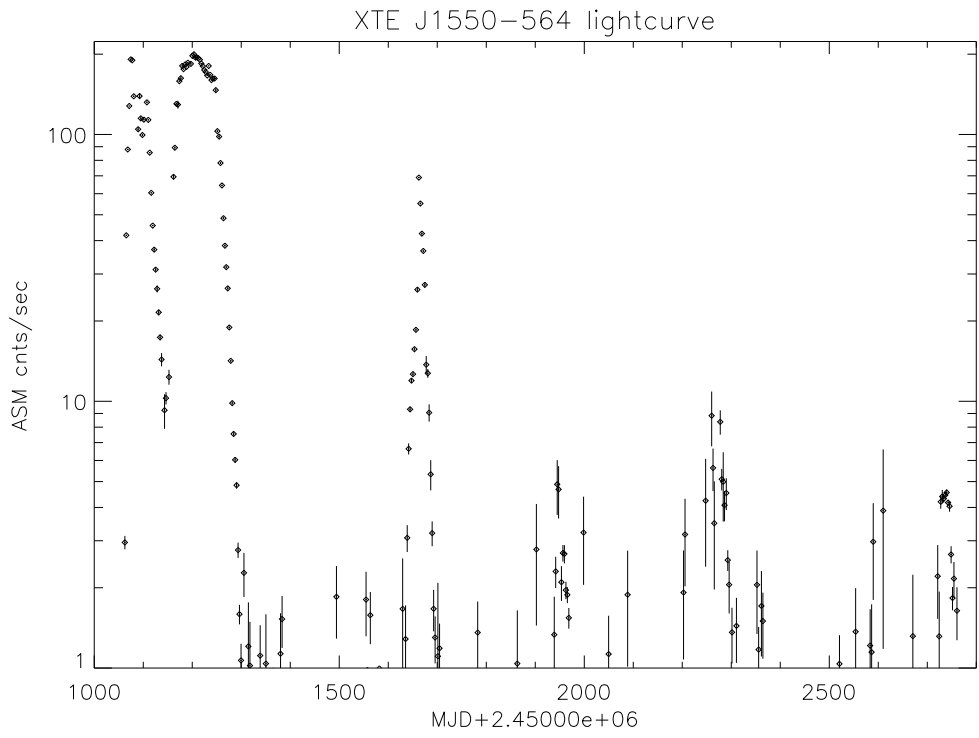}

\vfill

\caption{Long term lightcurve of XTE~J1550-564 in the 2-12 keV energy band 
according to the ASM/RXTE data.}
\end{figure*}
\pagebreak

\begin{figure*}[t]
\includegraphics[width=14cm]{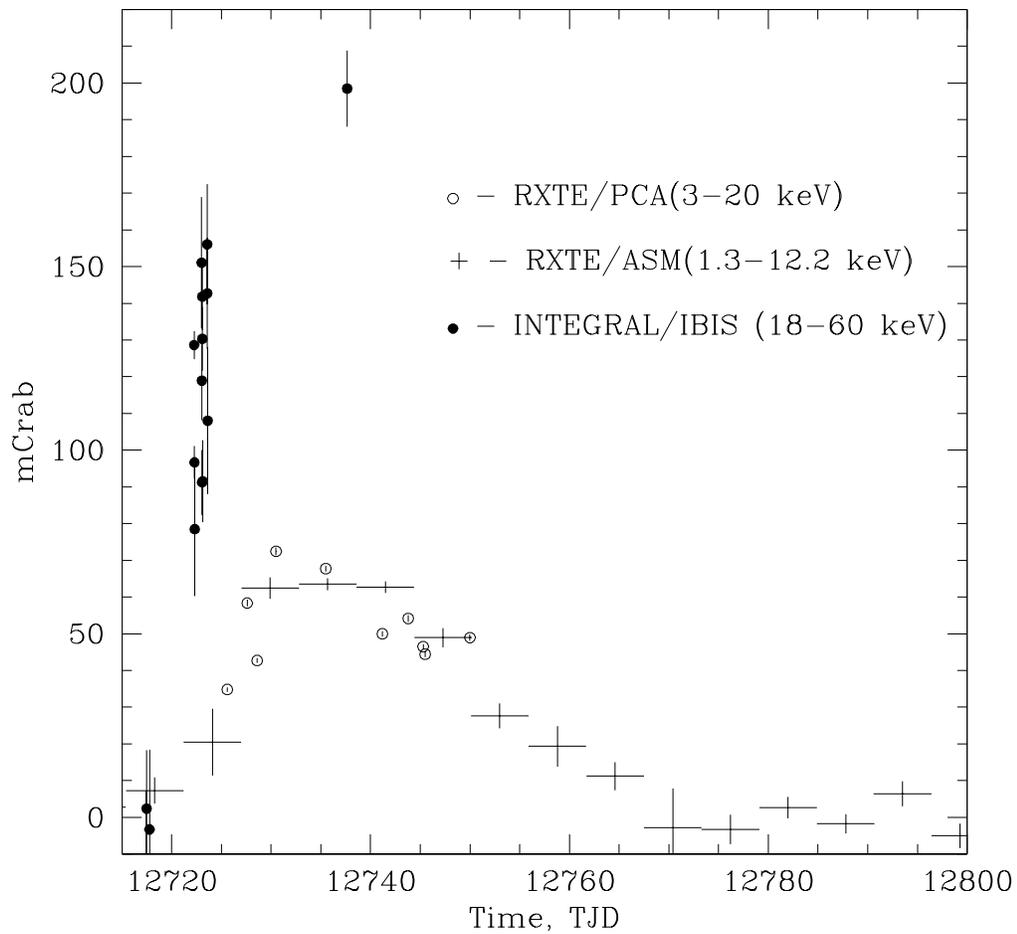}

\vfill

\caption{RXTE and INTEGRAL data of the XTE~J1550-564 lightcurve obtained
  during 2003 outburst.} 
\end{figure*}
\pagebreak

\begin{figure*}[t]
\includegraphics[width=14cm]{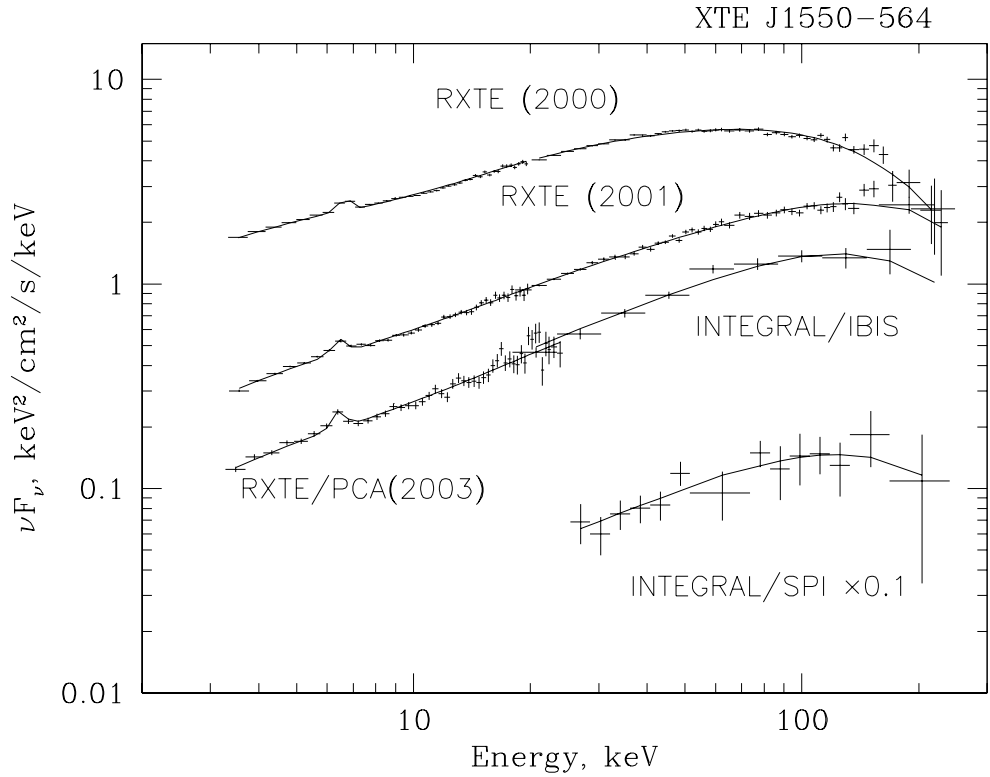}

\vfill

\caption{INTEGRAL and RXTE broadband spectra of XTE~J1550-564 obtained
  during 2003 outburst. Source spectra collected during hard state of
  2000 and during 2001 outbursts are shown for comparison.}
\end{figure*}

\end{document}